\RequirePackage{fix-cm}
\documentclass{article}
\usepackage{graphicx}
\usepackage{mathptmx}      
\usepackage{tikz}
\usepackage{tikz-cd}
\usepackage{algorithm}
\usepackage{algpseudocode}
\usepackage[colorlinks,linkcolor=black,anchorcolor=black,citecolor=black,urlcolor=blue]{hyperref}
\usepackage{mathtools}
\usepackage{amssymb}
\usepackage{mciteplus}
\usepackage[superscript]{cite}
\usepackage{soul}
\usepackage{caption}
\usepackage{subcaption}
\usepackage{float}
\usepackage{booktabs}

\usepackage{geometry}
\usepackage[utf8]{inputenc}
\usepackage{algpseudocode}
\usepackage{algorithm}
\usepackage{hyperref}
\usepackage{mathrsfs}
\hypersetup{
    colorlinks=true,
    linkcolor=blue,
    filecolor=magenta,      
    urlcolor=magenta,
    citecolor=blue
    }

\title{Persistent Sheaf Laplacian Analysis of Protein Flexibility }
\author{Nicole Hayes$^1$, 
Xiaoqi Wei$^1$\footnote{Current address: Department of Mathematics, North Carolina State University, Raleigh, NC.
},~ Hongsong Feng$^1$,
Ekaterina Merkurjev$^{1,2}$\footnote{Corresponding author,
	Email:  merkurje@msu.edu} ~ and 
 Guo-Wei Wei$^{1,3,4}$\footnote{Corresponding author,
	Email:  weig@msu.edu}\\
$^1$ Department of Mathematics, \\
Michigan State University, MI 48824, USA.\\
$^2$ Department of Computational Mathematics, Science and Engineering\\
Michigan State University, MI 48824, USA.\\
$^3$ Department of Electrical and Computer Engineering,\\
Michigan State University, MI 48824, USA. \\
$^4$ Department of Biochemistry and Molecular Biology,\\
Michigan State University, MI 48824, USA. \\
}

\begin{document}

\date{}

\maketitle

\abstract{ 
Protein flexibility, measured by the B-factor or Debye-Waller factor, is essential for protein functions such as structural support, enzyme activity, cellular communication, and molecular transport. Theoretical analysis and prediction of protein flexibility are crucial for protein design, engineering, and drug discovery. In this work, we introduce the persistent sheaf Laplacian (PSL), an effective tool in topological data analysis, to model and analyze protein flexibility. By representing the local topology and geometry of protein atoms through the multiscale harmonic and non-harmonic spectra of PSLs, the proposed model effectively captures protein flexibility and provides accurate, robust predictions of protein B-factors. Our PSL model demonstrates an increase in accuracy of 32\% compared to the classical Gaussian network model (GNM) in predicting B-factors for a dataset of 364 proteins. Additionally, we construct a blind machine learning prediction method utilizing global and local protein features. Extensive computations and comparisons validate the effectiveness of the proposed PSL model for B-factor predictions. \\

{\it Keywords:} 
Protein flexibility, persistent sheaf Laplacians, topological data analysis, machine learning, B-factor prediction. 
}

\newpage
 
\section{Introduction}
 

 Proteins are pivotal to life, playing an essential role in many biological processes, including signaling, gene regulation, transcription, translation, interaction with a protein or substrate molecule, etc. \cite{petsko2004protein} They are composed of amino acids, which form polypeptide chains and fold into specific three-dimensional (3D) structures. There are four levels of protein structures: primary, secondary, tertiary, and quaternary. The primary structure is the linear sequence of amino acids, whereas the secondary structure refers to $\alpha$-helices and $\beta$-sheets due to hydrogen bonds and electrostatic interactions. The tertiary structure corresponds to the 3D shape of a single polypeptide chain, while the quaternary structure describes the global arrangement of multiple polypeptide chains into a functional complex \cite{branden2012introduction}. 

Proteins have various functions; most notably, some of the functions of proteins include catalyzing metabolic reactions (enzymes), providing structural support (e.g., collagen in connective tissues), facilitating cellular communication (e.g., receptors and signaling molecules), and transporting molecules (e.g., hemoglobin for oxygen transport). These functions originate from their 3D structures. In particular, protein structure flexibility is a vital characteristic of protein structure that is essential to protein functions \cite{radivojac2004protein}. Specifically, protein flexibility enables proteins to adapt to various shapes and conditions, which facilitate their interactions with other molecules, such as DNA, RNA, ions, co-factors, ligands, and other small molecules. Under physiological conditions, proteins undergo constant thermal fluctuation, which enables the proteins to bind substrates, catalyze reactions, and transmit signals. Enzymes, for example, exhibit an induced fit mechanism, where their active sites adapt complementary shapes to accommodate substrates, improving the catalytic efficiency. In a similar way, molecular motors, such as myosins and kinesins, utilize flexibility to enable directed movement during muscle contraction and intracellular transport. 

Protein flexibility can be measured by the B-factor, also known as the Debye-Waller factor, \textcolor{black}{which measures the attenuation of X-ray or neutron scattering due to thermal motion of atoms in protein crystallography. Specifically, the B-factor is defined according to the mean displacement of a scattering center in X-ray diffraction data \cite{sun2019, bramer2020atom}. The B-factor is used to describe the flexibility of atoms and/or amino acids within a protein structure, and it further provides valuable information about the protein's thermal motion, structural stability, activity, and other protein functions} \cite{yuan2005prediction}.

Protein flexibility has been intensively studied in computational biophysics in recent decades \cite{ma2005usefulness,vihinen1994accuracy,jacobs2001protein,camps2009flexserv}.  In addition to the thoroughly investigated flexibility of proteins involved in folding, folded proteins (i.e., proteins in their native conformations) are also flexible and, in fact, exhibit internal motion in neighborhoods of their native conformations \cite{McCammon1977, huber1983}. In a seminal work, McCammon et al. \cite{McCammon1977} investigated such local motion in a small folded globular protein using a molecular dynamics (MD) approach, demonstrating the fluid-like characteristics of the internal motions. However, analyzing the dynamics of a large protein would require simulations at time scales that are intractable for the MD approach \cite{xia2013multiscale}. Consequently, other methods have since emerged using a time-harmonic approximation \cite{park2013coarse} to the protein's potential energy function used in MD, resulting in time-independent techniques. Such methods include normal mode analysis (NMA) \cite{Tasumi1982, Brooks1983, Go1983, Levitt1985, park2013coarse} and elastic network models (ENMs) \cite{Atilgan2001,Bahar1998,Bahar1997,Hinsen1998,Li2002,Tama2001}.

Some of the most popular methods \cite{xia2013multiscale, opron2014fast, xia2015multiscale} for protein flexibility analysis include the Gaussian network model (GNM) \cite{flory1976statistical, Bahar1997, haliloglu1997} and anisotropic network model (ANM) \cite{Atilgan2001}, both of which are types of ENMs. The GNM approach treats the protein as a network, with the residues representing the junctions. B-factors are then approximated using the first few eigenvalues of the connectivity matrix, which correspond to the long-time dynamics of proteins that MD simulations are unable to capture \cite{Yang2008Coarse}. Moreover, multiple methods have emerged as modifications of the original GNM and ANM models, including generalized GNM (gGNM), multiscale GNM (mGNM), and multiscale ANM (mANM) \cite{xia2015multiscale}. Such methods attempt to improve the efficiency and accuracy of GNM and ANM. Due to their ability to capture multiscale information intrinsic to protein structures, mGNM and mANM models have been shown \cite{xia2015multiscale} to significantly improve B-factor predictions of proteins compared to the original GNM and ANM methods. 

Other algorithms, such as the flexibility-rigidity index (FRI) \cite{xia2013multiscale}, which relies on the theory of continuum elasticity with atomic rigidity (CEWAR), have also improved results for B-factor prediction over the original GNM method. The FRI is based on the assumption that protein functions depend solely upon the protein's structure and environment, and therefore it assesses flexibility and rigidity by analyzing the topological connectivity and geometric compactness of protein structures. A benefit of the flexibility-rigidity index is that it bypasses the Hamiltonian interaction matrix and matrix diagonalization. Consequently, the FRI has significantly reduced computational complexity compared to other algorithms for protein flexibility analysis. Additional modifications, including fast FRI (fFRI) \cite{opron2014fast}, anisotropic FRI (aFRI) \cite{opron2014fast}, and multiscale FRI (mFRI) \cite{opron2015}, have been developed to further improve the efficiency of FRI as well as its accuracy on structures that are difficult for the NMA, GNM, and FRI algorithms\cite{opron2015}.

Recently, many machine learning approaches have been developed for protein flexibility analysis. \textcolor{black}{For example, sequence-based predictions have been reported \cite{schlessinger2005protein,de2012predyflexy,vander2021medusa}, and other machine-learning-based predictions of protein flexibility have also been proposed} \cite{vander2021medusa,masters2023deep,song2024accurate}.  More  recently, a method that utilizes  both sequence information and structure information has been developed for protein B-factor prediction \cite{xu2024opus}.

In 2019, persistent topological Laplacians (PTLs) \cite{wang2020persistent, chen2021evolutionary} were first introduced to overcome certain drawbacks of persistent homology, a key technique used in topological data analysis (TDA) \cite{carlsson2009topology,edelsbrunner2008persistent}.  
Many PTLs have been proposed in the past few years, including the persistent combinatorial Laplacian, the persistent path Laplacian, the persistent sheaf Laplacian (PSL) \cite{wei2025persistent}, the persistent directed graph Laplacian, and the persistent hyperdigraph Laplacian  \cite{wei2025persistent2}. Most of these algorithms are global, offering the topological and geometric descriptions of all objects in their topological space. In other words, they generate information about the protein as a whole. However, for protein flexibility analysis, one must have a method to describe the local properties of individual atoms. \textcolor{black}{The PSL model serves such a function, as it allows the assignment of a specific weight at each node (or atom); thus, it provides local topological and geometric information in its spectra, making it suitable for protein flexibility analysis.}    

The aim of the present work is to demonstrate the utility of the PSL model for protein flexibility analysis via the prediction of protein B-factors. The remainder of this manuscript is organized in the following manner: all results of this work are given in Section \ref{sec:results}. Section \ref{sec:results-subsets} summarizes our results on protein subsets from the literature, and Section \ref{sec:results-individual} presents the performance of the PSL model on individual proteins that are challenging for the GNM. Section \ref{sec:blind} details the results for blind machine learning prediction using the PSL model. In Section \ref{sec:method}, we describe the algorithms used in this manuscript, including some background on persistent homology and cellular sheaves.

\section{Results}
\label{sec:results}

\begin{figure}
    \centering
    \includegraphics[width=\textwidth]{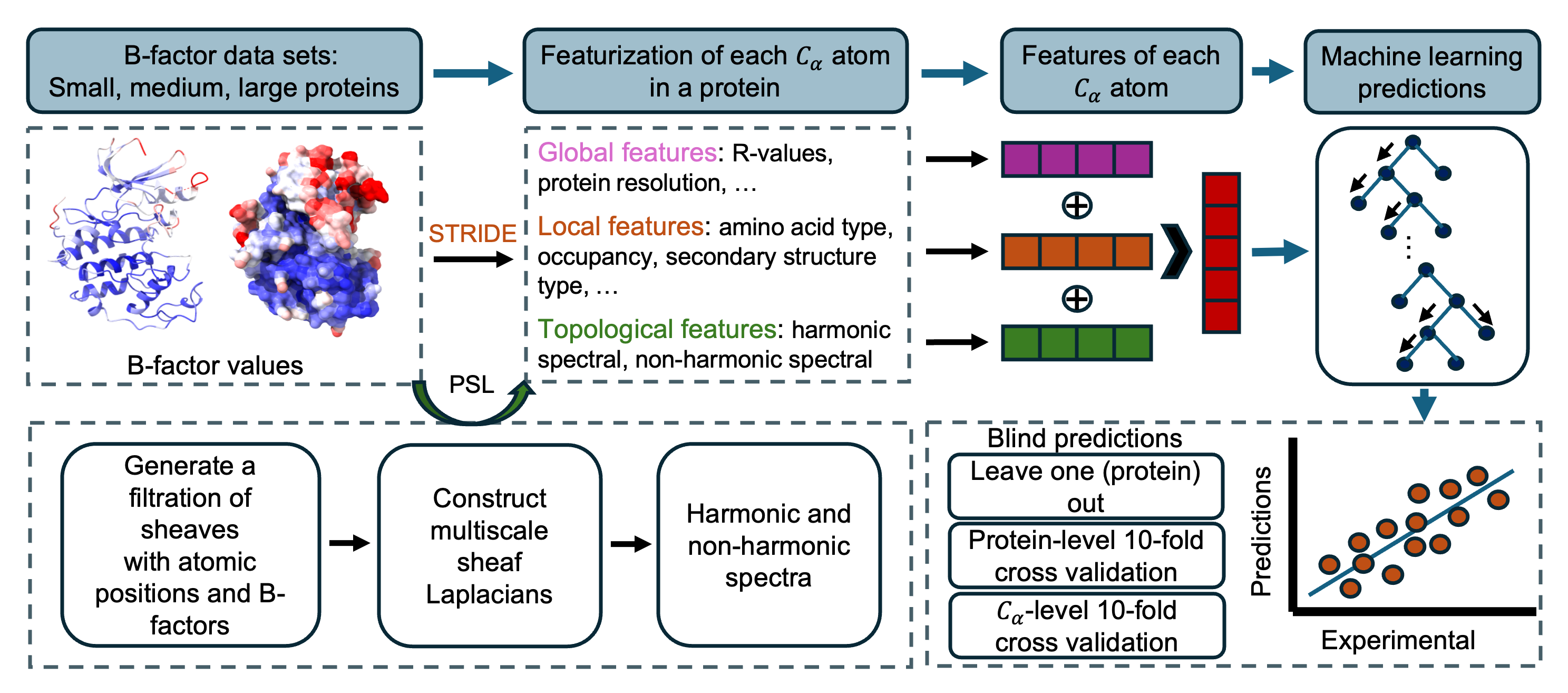}
    \caption{\textcolor{black}{Outline of the methods used in our work. The blind B-factor prediction in Section \ref{sec:blind} utilizes all pictured features, while the protein subset results from Section \ref{sec:results-subsets} include only the topological features generated using the persistent sheaf Laplacian (PSL) model.}}
    \label{concept-fig}
\end{figure}

\textcolor{black}{In this section, we present our results for experiments applying the persistent sheaf Laplacian (PSL) model as outlined in the previous section. Figure \ref{concept-fig} summarizes the methods used to generate the results throughout this section.}

\subsection{Results on protein subsets}
\label{sec:results-subsets}

\subsubsection{Data sets}

To demonstrate the persistent sheaf Laplacian model's performance on proteins of various sizes, we conducted computational experiments on four data sets. Three of these data sets were constructed by Park et al. \cite{park2013coarse} as sets of relatively small-, medium-, and large-sized protein structures. There are 33 proteins in the set of small-sized proteins, 36 in the set of medium-sized proteins, and 35 in the set of large-sized proteins. The fourth data set is a superset constructed by Opron et al. \cite{opron2014fast, opron2015} consisting of (1) the three aforementioned sets, (2) 40 proteins of varying sizes randomly selected from the Protein Data Bank (PDB) \cite{pdb1977}, and (3) 263 high-resolution protein structures used by Xia et al. \cite{xia2013multiscale} in tests of their FRI algorithm, with the duplicates subsequently removed. (Note that in their earlier paper, Opron et al. \cite{opron2014fast} used a set of 365 proteins, but their later manuscript \cite{opron2015} excluded the protein with PDB ID 1AGN due to an unrealistic B-factor. The present paper utilizes the updated set consisting of 364 proteins.)

Additionally, all protein data sets used for B-factor prediction in the present study were preprocessed to contain only the $C_{\alpha}$ atoms from their respective proteins. As discussed by Xia et al. \cite{xia2013multiscale}, the B-factor for an arbitrary atom in a protein is associated with that atom's flexibility, but its B-factor may be affected by diffraction in data collection, preventing a direct interpretation of flexibility. However, the B-factors of $C_{\alpha}$ atoms correlate directly with their atomic flexibility. Accordingly, our B-factor predictions in this work can be interpreted as atomic flexibility predictions.

Table \ref{subset-results} displays the results of the PSL model compared to other methods on the data sets of small, medium, and large proteins as well as the superset. 

\subsubsection{Parameters and results}

For all PSL results in this section \textcolor{black}{and Section \ref{sec:results-individual}}, we utilized a filtration induced by three radii: 6\r{A}, 9\r{A}, and 12\r{A}. For each radius, we generate a 0th persistent sheaf Laplacian matrix $L_0$ and compute its eigenvalues, then compute the maximum, minimum, mean, and median of the set of non-zero eigenvalues, as well as the number of zero eigenvalues. These quantities comprise five features for each radius, resulting in 15 features in total for each residue. To obtain the B-factor predictions \textcolor{black}{in this section}, we performed linear regression using the set of PSL features for the full set of 364 proteins \textcolor{black}{as well as the subsets}.

To better assess the performance of the PSL method relative to other approaches and to avoid overfitting, we did not perform an extensive search for the optimal filtration radii and eigenvalue statistic parameters for each task below. Rather, we conducted experiments on the set of 364 proteins with a few sets of parameters and chose those that yielded a good average Pearson correlation coefficient over the entire set. The above parameters may be tuned to further improve model performance for a given task---higher-order persistent sheaf Laplacian matrices and their respective eigenvalues may also be used to generate such features, and other statistics may be used as well, such as the standard deviation of the non-zero eigenvalues. Moreover, suitable filtration radii may be chosen to capture desired multiscale information for a given protein. Another example of PSL feature generation can be seen in Section \ref{sec:blind-features}.

\begin{table}[H]
    \centering
    \begin{tabular}{c|c|c|c|c|c|c|c}
    \hline
    Protein Set & PSL & ASPH (B) \cite{bramer2020atom} & ASPH (W) \cite{bramer2020atom} & opFRI \cite{opron2014fast} & pfFRI \cite{opron2014fast} & GNM \cite{park2013coarse} & NMA \cite{park2013coarse} \\
    \hline
    Small  & 0.927 & 0.85 & 0.86 & 0.667 & 0.594 & 0.541 & 0.480 \\
    Medium  & 0.728 & 0.69 & 0.69 & 0.664 & 0.605 & 0.550 & 0.482 \\
    Large  & 0.643 & 0.61 & 0.62 & 0.636 & 0.591 & 0.529 & 0.494 \\
    Superset & 0.751 & 0.65 & 0.66 & 0.673 & 0.626 & 0.565 & NA \\
    \hline
    \end{tabular}
\caption{Average Pearson correlation coefficients for the PSL model compared to other methods. Experiments were conducted on the full set of 364 proteins as well as three subsets of small, medium, and large protein structures as described by Park et al. \cite{park2013coarse} ASPH denotes the atom-specific persistent homology method developed by Bramer et al. \cite{bramer2020atom}, with results using Bottleneck (B) and Wasserstein (W) metrics displayed. Both sets of ASPH results used both an exponential and Lorenz kernel for least-squares fitting. opFRI and pfFRI results are from Opron et al. \cite{opron2014fast}, and GNM and NMA results are from Park et al. \cite{park2013coarse}. }
\label{subset-results}
\end{table}

 The PSL model achieves improved performance over all other compared methods on all data sets shown in Table \ref{subset-results}. In particular, the PSL model improves the benchmark GNM by 32\%.  

\subsection{Individual protein case studies}
\label{sec:results-individual}

As Opron et al. discussed in \textcolor{black}{their 2015 work} \cite{opron2015}, the Gaussian network model (GNM) experiences difficulty in predicting B-factors for certain protein structures. In addition to the comparison shown in Table \ref{subset-results}, in this section, we examine a few case studies of particular proteins to demonstrate the success of the PSL model on such structures. All protein structural visualizations were generated using the Visual Molecular Dynamics software (VMD) \cite{humphrey1996vmd}, and residues of each protein are assigned colors based on their experimental or predicted B-factors. Lower B-factors are shown as blue (corresponding to ``colder" or more rigid residues), and higher B-factors are shown as red (corresponding to ``warmer" or more flexible residues). All GNM results were obtained using the default GNM model with a cutoff of 7\r{A}.

\begin{figure}
    \centering
    \includegraphics[width=0.9\textwidth]{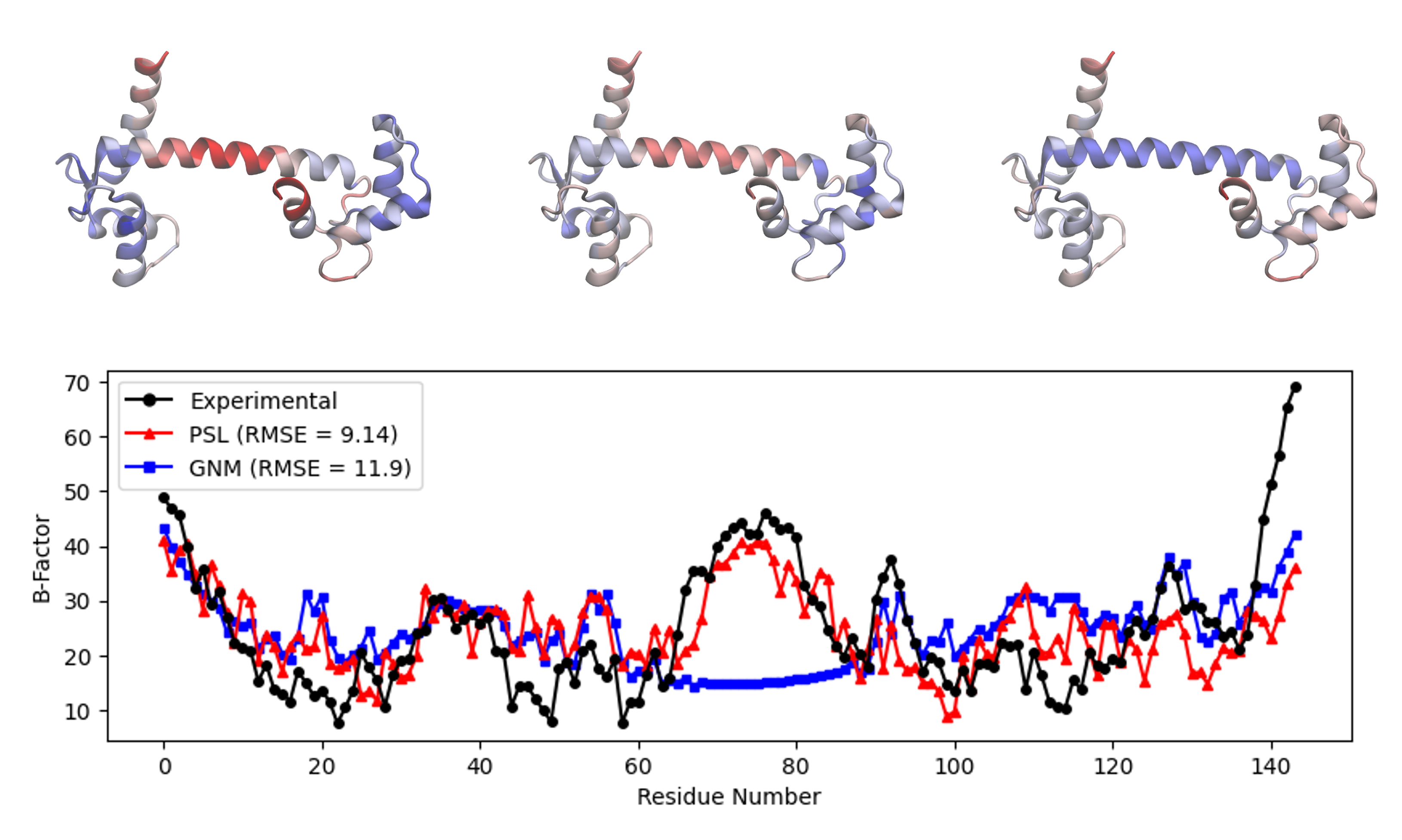}
    \caption{Top: visualization of the protein calmodulin (PDB ID: 1CLL) using Visual Molecular Dynamics (VMD) \cite{humphrey1996vmd}, with residues colored by experimental B-factors (left), B-factors predicted by PSL (center), and B-factors predicted by GNM (right). Bottom: experimental and predicted B-factors for each residue of the protein. The GNM result uses the default cutoff of 7\r{A}. The GNM underestimates the B-factors for residues between about 65 and 85.}
    \label{1cll-comparisons-fig}
\end{figure}

Calmodulin is a calcium detector within the cells and plays a significant role in numerous cellular pathways. Its flexibility allows it to  interact with varied target proteins. 
  Figure \ref{1cll-comparisons-fig} displays the predicted and experimental B-factors for the calcium-binding protein calmodulin (PDB ID: 1CLL) \cite{pdb1977} using our persistent sheaf Laplacian model as well as the Gaussian network model. We observe that the Gaussian network model produces a large error in B-factor prediction for residues from about 65-85. These residues correspond to a flexible hinge region of the protein \cite{opron2015}. \textcolor{black}{The root mean square error (RMSE) for the PSL model is 9.14 for calmodulin, a 23\% decrease from the GNM model's RMSE of 11.9.}

\begin{figure}
    \centering
    \includegraphics[width=0.9\textwidth]{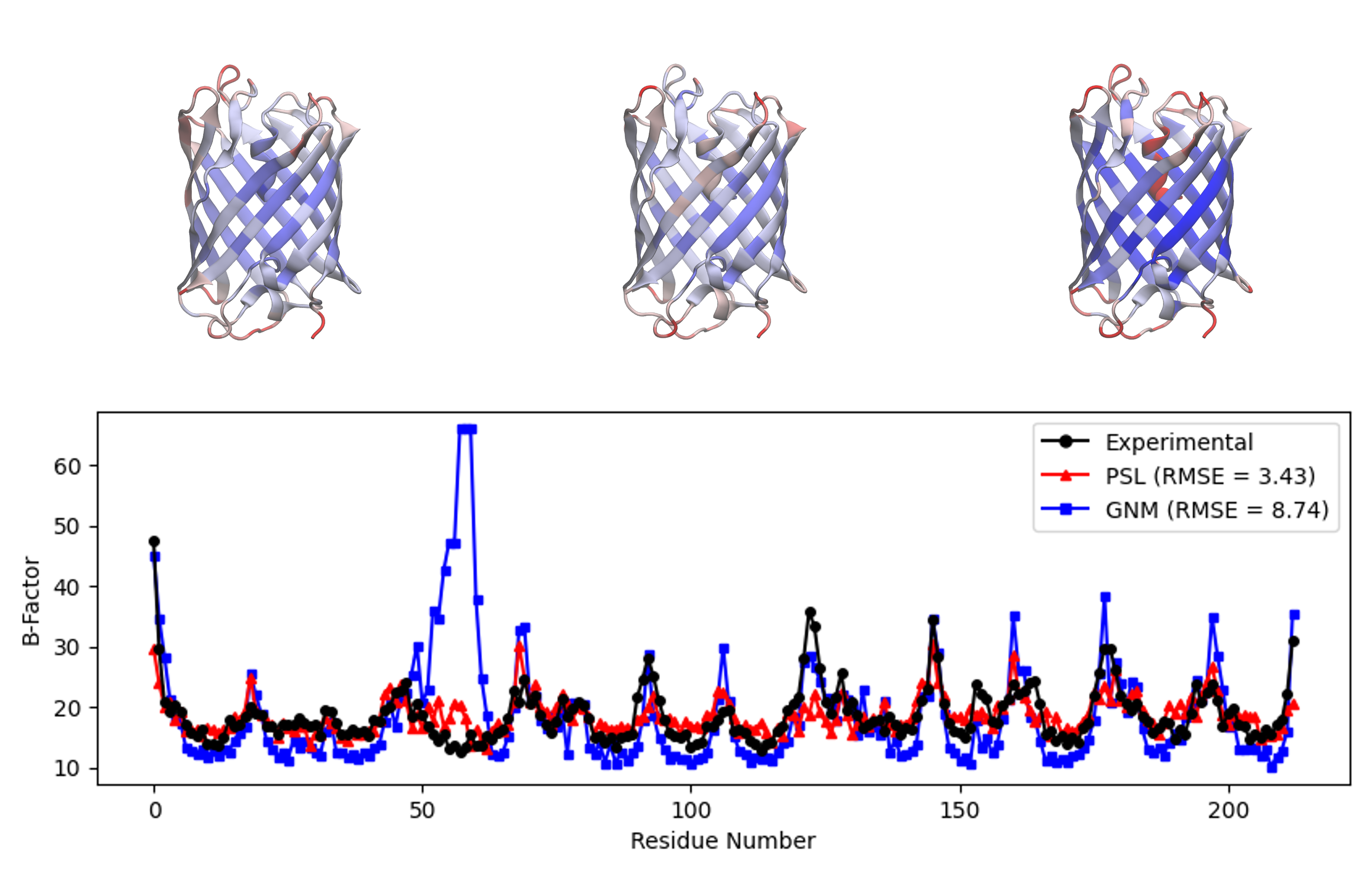}
    \caption{Top: visualization of the protein mTFP1 (PDB ID: 2HQK) using VMD \cite{humphrey1996vmd}, with residues colored by experimental B-factors (left), B-factors predicted by PSL (center), and B-factors predicted by GNM (right). Bottom: experimental and predicted B-factors for each residue of the protein. The GNM result uses the default cutoff of 7\r{A}. The GNM vastly overestimates the B-factors of residues around 50-60.}
    \label{2hqk-comparisons-fig}
\end{figure}

Next, we consider a monomeric cyan fluorescent protein (mTFP) that emits cyan light. It is used in biological experiments to visualize specific targets. Figure \ref{2hqk-comparisons-fig} shows experimental B-factors and predicted B-factors of the protein mTFP1 (PDB ID: 2HQK). Again, the predicted B-factors shown were computed using the Gaussian network model and our PSL model. As in the results for the protein calmodulin, the GNM is unable to correctly predict B-factors for one range of residues (around residues 50-60) in the protein mTFP1. Here, however, the Gaussian network model overestimates the B-factors in this region, visible in the GNM structural representation as the red $\alpha$-helix in the center of the $\beta$-barrel \cite{opron2015}. Opron et al. \cite{opron2015} observed that using a cutoff of 8\r{A} for GNM somewhat resolves this error, and they suggested that the GNM may experience difficulty in this region due to its use of hard thresholds based on connectivity parameters. The persistent sheaf Laplacian model is significantly more accurate in this region, likely due to the fact that it captures atom-specific information as well as molecular information at multiple scales. \textcolor{black}{Overall, the PSL model improves the RMSE on mTFP1 to 3.43 from 8.74 for the GNM, a nearly 61\% decrease.}

\begin{figure}
    \centering
    \includegraphics[width=0.9\textwidth]{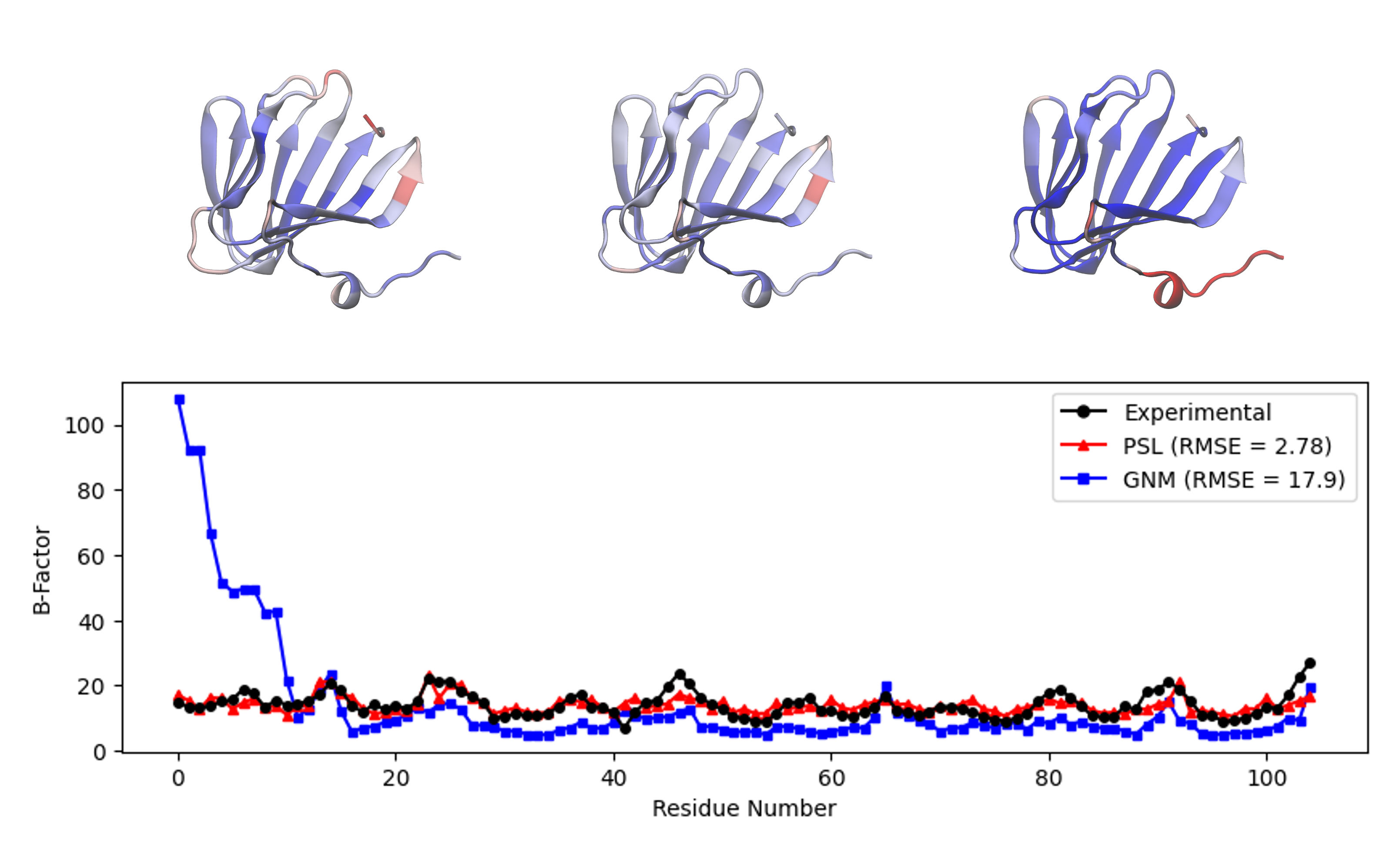}
    \caption{Top: visualization of the protein with PDB ID 1V70 using VMD \cite{humphrey1996vmd}, with residues colored by experimental B-factors (left), B-factors predicted by PSL (center), and B-factors predicted by GNM (right). Bottom: experimental and predicted B-factors for each residue of the protein. The GNM result uses a cutoff of 7\r{A}. The GNM vastly overestimates the B-factors for residues from about 0-10.}
    \label{1v70-comparisons-fig}
\end{figure}

We further consider a probable antibiotics synthesis protein from Thermus thermophilus. 
In Figure \ref{1v70-comparisons-fig}, we investigate the experimental and predicted B-factors of this protein (PDB ID: 1V70). On this protein, our persistent sheaf Laplacian model is able to predict the B-factors accurately across all residues of the protein, while the Gaussian network model experiences a high level of inaccuracy on residues from about 0-10. \textcolor{black}{This vast overprediction contributes to a very high RMSE value for the GNM, at 17.9. Our PSL model achieves a significantly lower RMSE of 2.78 on the protein 1V70, 84\% lower than that of the GNM.}

\begin{figure}
    \centering
    \includegraphics[width=0.9\textwidth]{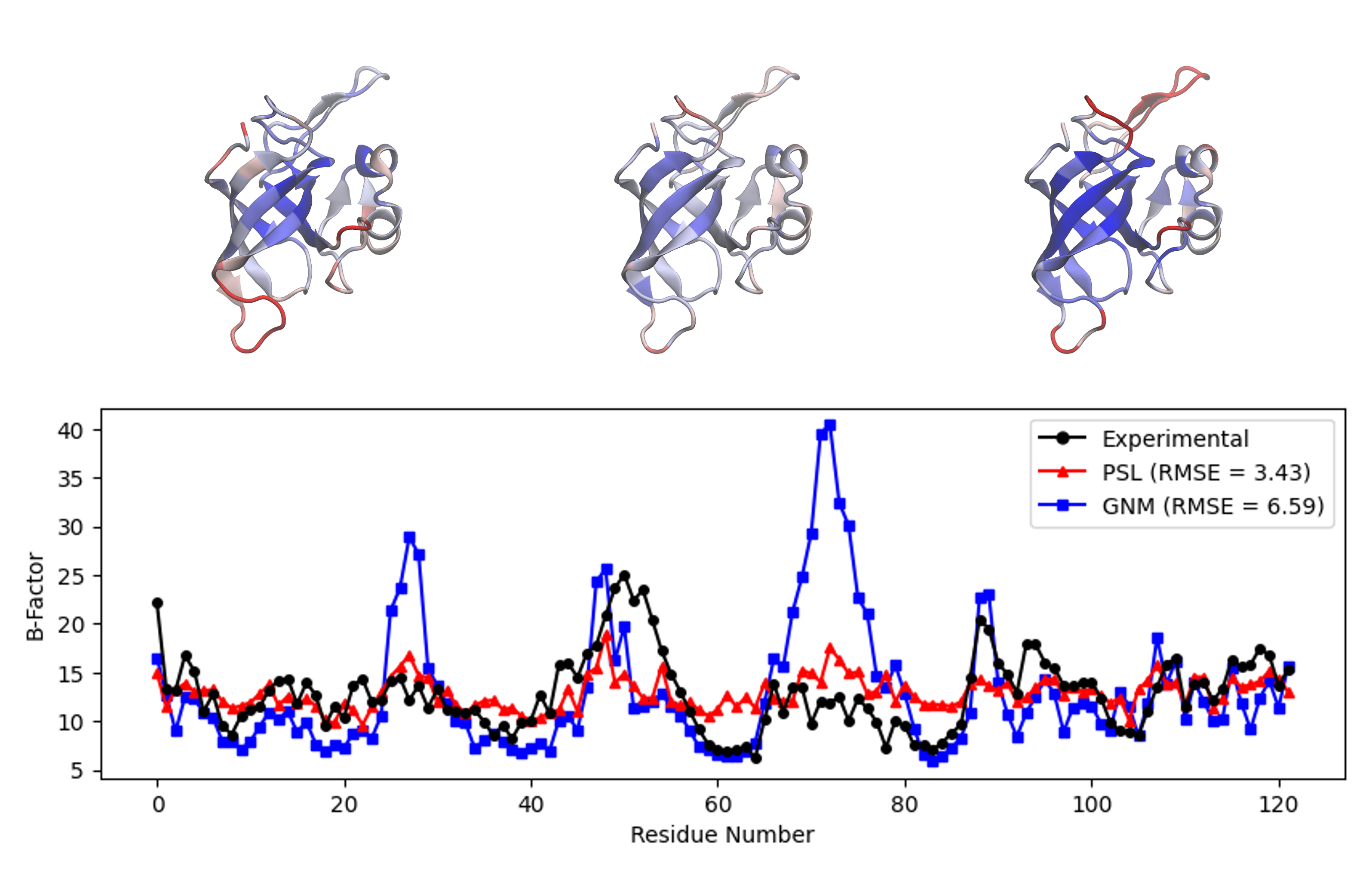}
    \caption{Top: visualization of the ribosomal protein L14 (PDB ID: 1WHI) using VMD \cite{humphrey1996vmd}, with residues colored by experimental B-factors (left), B-factors predicted by PSL (center), and B-factors predicted by GNM (right). Bottom: experimental and predicted B-factors for each residue of the protein. The GNM result uses a cutoff of 7\r{A}. The GNM overestimates the B-factors for residues between 60-80.}
    \label{1whi-comparisons-fig}
\end{figure}

Finally, we studied the ribosomal protein L14 (PDB ID: 1WHI) \cite{opron2015}, one of the most conserved ribosomal proteins. It functions as an organizational component of the translational apparatus. 
In Figure \ref{1whi-comparisons-fig}, we show the experimental and predicted B-factors for the ribosomal protein L14. Again, we observe that the GNM overestimates the flexibility of some regions of this protein, most significantly for the residues around 60-80. \textcolor{black}{The RMSE for the PSL model on this protein is nearly half that of the GNM model, whose RMSE is 6.59.}

\subsection{Blind machine learning prediction}\label{sec:blind}

\subsubsection{Data sets}

Two datasets, one from \textcolor{black}{Opron et al.}~\cite{opron2014fast,opron2015} and the other from \textcolor{black}{Park et al.}~\cite{park2013coarse} are used in our work. The first dataset contains 364 proteins~\cite{opron2014fast,opron2015}, and the second \cite{park2013coarse} has three sets of proteins with small, medium, and large sizes, which are the subsets of the 364 protein set. 

In our blind predictions, proteins 1OB4, 1OB7, 2OXL, and 3MD5 from the superset are excluded  because the STRIDE software cannot generate features for these proteins. We exclude protein 1AGN due to the known problems with this protein data~\cite{opron2014fast,opron2015}.  Additional proteins from the superset are also excluded. Proteins 1NKO, 2OCT, and 3FVA are excluded because these proteins have unphysical  B-factors (i.e., zero values). We also excluded proteins 3DWV, 3MGN, 4DPZ, 2J32, 3MEA, 3A0M, 3IVV, 3W4Q, 3P6J, and 2DKO due to inconsistent protein data processed with STRIDE compared to original PDB data. A total of 346 proteins are used for blind predictions. Those data can be found in our provided GitHub repository.

\subsubsection{PSL features}\label{sec:blind-features}

The second approach to B-factor prediction that we examined is a blind prediction for protein B-factors. We use PSL features as local descriptors of protein structures, applying three cutoff distances, i.e., 7, 10, and 13\AA, to define the atom groups used to construct a sheaf Laplacian matrix. For each cutoff distance, we generate a sheaf Laplacian matrix, 
$L_1$,  with a filtration radius matching the cutoff distance. From each matrix, we extract five features: the count of zero eigenvalues, and the maximum, minimum, mean, and standard deviation of the non-zero eigenvalues. Together, these provide 15 PSL features for blind machine learning predictions.

\subsubsection{Additional features}

In addition to PSL features, we extract a range of global and local protein features for building machine learning models. Each PDB structure is associated with global features, such as the R-value, resolution, and the number of heavy atoms, which are extracted from the PDB files. These features enable the comparison of the B-factors in different proteins.  
The local characteristics of each protein consist of packing density, amino acid type, occupancy, and secondary structure information generated by STRIDE \cite{heinig2004stride}. STRIDE provides comprehensive secondary structure details for a protein based on its atomic coordinates from a PDB file, classifying each atom into categories such as $\alpha$-helix, 3-10-helix, $\pi$-helix, extended conformation, isolated bridge, turn, or coil. Furthermore, STRIDE provides $\phi$ and $\psi$ angles and residue solvent-accessible area, contributing a total of 12 secondary features. In our implementation, we use one-hot encoding for both amino acid types and the 12 secondary features. The packing density of each $C_{\alpha}$ atom in a protein is calculated based on the density of surrounding atoms, with short, medium, and long-range packing density features defined for each $C_{\alpha}$ atom. The packing density of the $i$th $C_{\alpha}$ atom is defined as
\begin{align}\
	p_{i}^d =\frac{N_d}{N},
\end{align}	
where $d$ represents the specified cutoff distance in \AA, $N_d$ denotes the number of atoms within the Euclidean distance $d$ from the $i$th atom, and $N$ is the total number of heavy atoms in the protein. The packing density cutoff values used in this study are provided in Table \ref{table:packing-density}.
\begin{table}[htb!]
	\centering
	\begin{tabular}{c | c | c  }
		\hline
		Short &  Medium & Long\\
		\hline
		$d<3$ & $3\ge d<5$ & $5\le d$\\
		\hline
	\end{tabular}
	\caption{Packing density parameter in distance $d \AA$.}
	\label{table:packing-density}
\end{table}
Our PSL features, combined with the global and local features provided for each PDB file, offer a comprehensive feature set for each $C_{\alpha}$ atom in the protein. For blind predictions, we integrate these features with machine learning algorithms to build regression models. To evaluate the performance of our machine learning model on blind predictions, we conducted two validation tasks: 10-fold cross-validation and leave-one-(protein)-out validation. For 10-fold cross-validation, we designed two types of experiments---one based on splitting by PDB files and another on splitting by all $C_{\alpha}$ atoms collected from the PDB files. Our modeling and predictions are centered on the B-factors of $C_{\alpha}$ atoms. 


\subsubsection{Evaluation metrics}
To  assess our method for B-factor prediction, we use  the Pearson correlation coefficient (PCC):
\begin{align*}
	\text{PCC}({\bf{x}},{\bf{y}})=\frac{\sum_{m=1}^{M}(B_m^e-\bar{B}^e)(B_m^t-\bar{B}^t)}{\sqrt{\sum_{m=1}^{M} (B_m-\bar{B}^e)^2\sum_{m=1}^{M} (B_m^t-\bar{B}^t)^2}},
\end{align*}  
where $B_m^t, m = 1,2,\cdots,N$ are the predicted B-factors and $B_m^e, m = 1,2,\cdots,N$ are the experimental B-factors from the PDB file. Here $\bar{B}^e$ and $\bar{B}^t$ are the averaged B-factors.


\subsubsection{Machine learning algorithms}

For the blind predictions, instead of using more sophisticated methods \cite{merkurjev2020fast,mbo,merkurjev_aml}, we consider two simple machine learning algorithms, namely gradient-boosting decision trees (GBDT) and random forests (RF), to highlight the proposed PSL method. The hyperparameters of these two types of algorithms are given in Table \ref{table:rf-gbdt-parameters}.
\begin{table}
	\centering
	\small
	\begin{tabular}{c |c }		
		\toprule
		RF parameters &  GBDT parameters  \\ 
		\hline
		& $\rm n{\_}estimators = 1000$  	 \\
		$\rm n{\_}estimators = 1000$&$\rm max{\_}depth = 7$ 	\\
		$\rm max{\_}depth = 8$ & $\rm min{\_}samples{\_}split = 5$  \\
		$\rm min{\_}samples{\_}split = 4$&$ \rm subsample  = 0.8$  	\\
		$ \rm min{\_}samples{\_}leaf  = 0.8$&$\rm learning{\_}rate = 0.002$ \\		
		& $\rm max{\_}features = "sqrt"$  	\\
		\bottomrule
	\end{tabular}
	\caption{Hyperparameters of the random forest (RF) and gradient boosting decision tree (GBDT) algorithms used for the B-factor predictions.}
	\label{table:rf-gbdt-parameters}
\end{table}


\subsubsection{Machine learning results}
 
We carried out several experiments, the first of which is a leave-one-(protein)-out prediction using the four datasets described above. We trained models five times independently with different random seeds and calculated the average Pearson correlation coefficients from the ten sets of modeling predictions. Our results are shown in Table \ref{table:Pearson-values-LOO}, where the GBDT-based models yield better predictions than the RF-based models, as expected.
\begin{table}[htb!]
	\centering
	\begin{tabular}{c | c c  c c }
		\hline
		\textbf{Protein set} &  \textbf{RF} & \textbf{GBDT}\\
		\hline
		Small & 0.478& 0.433\\
		Medium& 0.518&0.590\\
		Large & 0.508& 0.582\\
		Superset & 0.542 &0.588\\
		\hline
	\end{tabular}
	\caption{Average Pearson correlation coefficients (PCC) of leave-one (protein)-out predictions for the four B-factor datasets. The PCC results obtained with random forest  (RF) and gradient boosting decision tree (GBDT) models are compared.}
	\label{table:Pearson-values-LOO}
\end{table}

In our study, we additionally carried out 10-fold cross-validation at the protein level. In each fold, we use nine out of the ten subsets of the 346 proteins to train our model, while the remaining subset is reserved for testing. Specifically, features of C$_{\alpha}$ atoms in the training proteins are pooled together to train the models, while those in the test proteins are used for evaluation. This process is repeated across ten different splits. Table \ref{table:Pearson-values-CV10-proteinwise} shows the average PCC values for two types of machine learning models. Again, the GBDT model gives better predictions than the RF model. 

\begin{table}[htb!]
	\centering
	\begin{tabular}{c | c c  c c }
		\hline
		\textbf{Protein set} &  \textbf{RF} & \textbf{GBDT}\\
		\hline
		 Superset & 0.397 & 0.452\\ 
		\hline
	\end{tabular}
	\caption{Average Pearson correlation coefficient (PCC) from protein-level 10-fold cross validation predictions with the collected 346 proteins. The B-factor values of $C_{\alpha}$ atoms in each protein are predicted. The average PCC value is calculated from five independent experiments. The PCC results with random forest (RF) and gradient boosting decision tree (GBDT) modeling are compared.}
	\label{table:Pearson-values-CV10-proteinwise}
\end{table}

We also performed an alternative C$_\alpha$-level 10-fold cross-validation. The dataset consists of more than 74,000 C$_\alpha$ atoms from 364 proteins.  In each of ten independent models, nine out of ten subsets of C$_\alpha$ atoms are used to train the models, while the remaining subset is used for testing.  As shown in Table \ref{table:Pearson-values-CV10-atomwise}, GBDT modeling yields slightly better predictions than RF-based modeling. 

\begin{table}[htb!]
	\centering
	\begin{tabular}{c | c c  c c }
		\hline
		\textbf{Protein set} &  \textbf{RF} & \textbf{GBDT}\\
		\hline
            Superset & 0.839 & 0.840\\
		\hline
	\end{tabular}
	\caption{Average Pearson correlation coefficient (PCC) from C$_\alpha$-level   10-fold cross validation predictions with all C$_\alpha$ atoms in the collected 346 proteins. The average PCC value is calculated from five independent experiments. The PCC results with random forest (RF)  and gradient boosting decision tree (GBDT) models are compared.}
	\label{table:Pearson-values-CV10-atomwise}
\end{table}

\section{Methods}
\label{sec:method}

\subsection{Persistent homology and persistent Laplacians}
As one of the most abstract mathematical subjects, homology excessively simplifies complex geometry. In contrast, persistent homology balances simplification and information retrieval in data analysis and is widely used in topological data analysis \cite{carlsson2009topology, edelsbrunner2008persistent}. However, persistent homology has several drawbacks, including its insensitivity to homotopic shape evolution. To address this challenge, the persistent spectral graph, also known as persistent Laplacians, was introduced on simplicial complexes in 2019 \cite{wang2020persistent}. Since then, various persistent Laplacians, or persistent topological Laplacians, have been proposed for different topological objects, such as path complexes, directed flag complexes, hyperdigraphs, and cellular sheaves \cite{wei2025persistent2}. 
      
Given a finite set $V$, a simplicial complex $X$ is a collection of subsets of $V$, 
such that if a set $\sigma$ is in $X$, then any subset of $\sigma$ is also in $X$. A set $\sigma$ that consists of $q+1$ elements is referred to as a $q$-simplex. If $\sigma$ is a subset of $\tau$, then we say that $\sigma$ is a face of $\tau$ and denote the face relation by $\sigma \leqslant \tau$.
If $X$ and $Y$ are simplicial complexes and $X\subset Y$, then $X$ is referred to as a subcomplex of $Y$. A simplicial complex $X$ gives rise to a simplicial chain complex 
\[
    \begin{tikzcd}[column sep = large]
        \centering
    \cdots \arrow[r, "\partial_{3}"] & 
    C_{2}(X) \arrow[r, "\partial_{2}"] & 
    C_{1}(X)  \arrow[r, "\partial_1"] & 
    C_{0}(X) \arrow[r] & 
    0.
    \end{tikzcd} 
\]
The real vector space $C_q(X)$ is generated by $q$-simplices. An element of $C_q(X)$ is called a $q$-chain.
The boundary operator $\partial_q$ is a linear map defined by 
\begin{align*}
    \partial_q [v_{a_0}, \dots, v_{a_q}] = \sum_i (-1)^i[v_{a_0}, \dots, \hat{v}_{a_i}, \dots, v_{a_q}],
\end{align*}
where the symbol $\hat{v}_{a_i}$ means that $\hat{v}_{a_i}$ is deleted. The total ordering of $V$ ensures that the boundary operator is well-defined. The $q$-th homology group $H_q=\ker \partial_{q}/\text{im} \partial_{q+1}$ is well defined since $\partial^2=0$. 
Now suppose $X$ is a subcomplex of $Y$. Then we have the following diagram
\[
    \begin{tikzcd}
        \cdots \arrow[r, "\partial_{q+2}^X"]
        & C_{q+1}(X) \arrow[r, "\partial_{q+1}^X"] \arrow[d, hook, dashed] 
          & C_{q}(X) \arrow[r, "\partial_q^X"] \arrow[d, hook, dashed] 
            & C_{q-1}(X) \arrow[r, "\partial_{q-1}^X"] \arrow[d, dashed, hook]
              & \cdots
            \\
        \cdots \arrow[r, "\partial_{q+2}^Y"]
        & C_{q+1}(Y) \arrow[r, "\partial_{q+1}^Y"] 
          & C_q(Y) \arrow[r, "\partial_q^Y"] 
            & C_{q-1}(Y) \arrow[r, "\partial_{q-1}^Y"]
              & \cdots 
    \end{tikzcd}
\]where hooked dashed arrows represent inclusion maps $\iota: C_q(X) \hookrightarrow C_q(Y)$. The inclusion $\iota$ induces a map $\iota^{\bullet}: H_q(X) \to H_q(Y)$. 
The $q$-th persistent homology for the pair $(X, Y)$ is the image
\begin{align*}
    \iota^{\bullet} (H_q(X)).
\end{align*}
Usually the ranks of persistent homology groups are represented by barcodes, where each bar represents a topological feature that persists in the filtration, offering a multiscale topological characterization of the input point cloud\cite{carlsson2009topology, edelsbrunner2008persistent}.

Recently, the theory of persistent Laplacians \cite{ wang2020persistent} has been proposed to extract additional information from a point cloud. A persistent Laplacian is a positive semi-definite operator whose kernel is isomorphic to the corresponding persistent homology group. The additional information provided by the non-zero eigenvalues of persistent Laplacians can be learned by machine learning algorithms.
Since $C_q(X)$ is generated by $q$-simplices, it is equipped with a canonical inner product. 
Let $C_{q+1}^{X,Y}=\{c \in C_{q+1}(Y) \mid \partial_{q+1}^Y(c) \in C_q(X) \}$ 
and $\partial_{q+1}^{X,Y}$ the restriction of $\partial^Y_{q+1}$ to $C_{q+1}^{X,Y}$.
The $q$-th persistent Laplacian $\Delta_{q}^{X,Y}$ is defined by 
\begin{align}
    \partial_{q+1}^{X,Y} (\partial_{q+1}^{X,Y})^{\dag} + (\partial^X_{q})^{\dag} \partial^X_{q},
\end{align}
where $\dag$ denotes the adjoint of a linear morphism. Using basic linear algebra we can prove that the kernel of $\Delta_{q}^{X,Y}$ is isomorphic to $\iota^{\bullet}(H_q(X))$. 
Generally speaking, any method that utilizes multiscale Laplacians to analyze data can be referred to as a persistent Laplacian method.

\subsection{Cellular sheaves and persistent sheaf Laplacians}

Molecular structures often contain important non-spatial information, and many applications of topological methods in analyzing molecular data require integration of non-spatial information. For example, we can use generalized distance to model the biochemical interaction between atoms or only use specific types of atoms as input to persistent homology \cite{cang2018integration} or persistent Laplacians \cite{wang2020persistent}. An alternative approach is to integrate biological information through the construction of (co)chain complexes and extend persistent homology and persistent Laplacians to new settings. For example, one can construct a filtration of cellular sheaves and consider the persistence module of sheaf cochain complexes instead of simplicial complexes and simplicial chain complexes\cite{hansen2019learning}.

Roughly speaking, a cellular sheaf $\mathscr{F}$ is a simplicial complex $X$ with 
an assignment to each simplex $\sigma$ of $X$ a finite-dimensional vector space $\mathscr{S}(\sigma)$ (referred to as the stalk of $\mathscr{S}$ over $\sigma$)
and to each face relation $\sigma \leqslant \tau$ (i.e.,\ $\sigma \subset \tau$)
a linear morphism of vector spaces denoted by
$\mathscr{S}_{\sigma \leqslant \tau}$ (referred to as the restriction map of the face relation $\sigma \leqslant \tau$), satisfying the rule 
\begin{align*}
    \rho \leqslant \sigma \leqslant \tau \Rightarrow \mathscr{S}_{\rho \leqslant \tau} = \mathscr{S}_{\sigma \leqslant \tau} \mathscr{S}_{\rho \leqslant \sigma}
\end{align*}
and $\mathscr{S}_{\sigma \leqslant \sigma}$ is the identity map of $\mathscr{S}(\sigma)$. We can view stalks as information stored for each simplex, and restriction maps as the way this information interacts.
A cellular sheaf gives rise to a sheaf cochain complex
\[
    \begin{tikzcd}
        \centering
    0 \arrow[r] & 
    C^0(X; \mathscr{S}) \arrow[r, "d"] & 
    C^1(X; \mathscr{S}) \arrow[r, "d"] & 
    C^2(X; \mathscr{S}) \arrow[r, "d"] & 
    \cdots.
    \end{tikzcd} 
\]
The $q$-th sheaf cochain group $C^q(X; \mathscr{S})$ is the direct sum of stalks over $q$-dimensional simplices.
To define coboundary maps $d$, we can globally orient the simplicial complex $X$ and obtain a signed incidence relation, an assignment to each $\sigma \leqslant \tau$ an integer $[\sigma: \tau]$. 
The coboundary map $d^{q}: C^q(X;\mathscr{S}) \to C^{q+1}(X;\mathscr{S})$ is defined by 
\begin{align*}
    d^q \vert_{\mathscr{S}(\sigma)} = \sum_{\sigma \leqslant \tau} [\sigma: \tau] \mathscr{S}_{\sigma \leqslant \tau}.
\end{align*} 
Now suppose we have $\mathscr{F}$ on $X$ and $\mathscr{G}$ on $Y$ such that 
$X \subseteq Y$ and stalks and restriction maps of $X$ are identical to those of $Y$. If each stalk is an inner product space then
we have the following diagram
\[
\begin{tikzcd}
    C^{q-1}(X; \mathscr{F}) \arrow[rr, "d^{q-1}_{\mathscr{F}}", shift left] 
     &
      & C^{q}(X; \mathscr{F}) \arrow[ll, "(d^{q-1}_{\mathscr{F}})^{\dag}", shift left] \arrow[rd, "d^{q}_{\mathscr{F}, \mathscr{G}}", shift left] \arrow[dd, hook, dashed]
       & 
        & \\
     & 
      &
       & \Theta^{q+1}_{\mathscr{F}, \mathscr{G}} \arrow[lu, "(d^{q}_{\mathscr{F}, \mathscr{G}})^{\dag}", shift left] \arrow[rd, hook, dashed]
        & \\
     &
      & C^q(Y; \mathscr{G}) \arrow[rr, "d^q_{\mathscr{G}}", shift left]
       & 
        & C^{q+1}(Y; \mathscr{G}) \arrow[ll, "(d^q_{\mathscr{G}})^{\dag}", shift left]
\end{tikzcd}
\]
where $\Theta_{\mathscr{F}, \mathscr{G}}^{q+1} = \{x \in C^{q+1}(Y; \mathscr{G}) \mid (d^{q}_{\mathscr{G}})^{\dag}(x) \in C^{q}(X; \mathscr{F})\}$ and 
$d^{q}_{\mathscr{F}, \mathscr{G}}$ is the adjoint of $\pi (d^{q}_{\mathscr{G}})^{\dag}\vert_{\Theta^{q+1}_{\mathscr{F}, \mathscr{G}}}: \Theta^{q+1}_{\mathscr{F}, \mathscr{G}} \to C^q(X;\mathscr{F})$ ($\pi$ is the projection map from $C^q(Y; \mathscr{G})$ to its subspace $C^q(X; \mathscr{F})$).
We define the $q$-th persistent sheaf Laplacian $\Delta_q^{\mathscr{F}, \mathscr{G}}$ by
\begin{align*}
    \Delta_q^{\mathscr{F}, \mathscr{G}} = (d^{q}_{\mathscr{F}, \mathscr{G}})^{\dag} d^{q}_{\mathscr{F}, \mathscr{G}} +  d^{q-1}_{\mathscr{F}}(d^{q-1}_{\mathscr{F}})^{\dag}.
\end{align*}
When $\mathscr{F}= \mathscr{G}$, the persistent sheaf Laplacian is equal to the sheaf Laplacian of $\mathscr{F}$. When $\mathscr{F}$ and $\mathscr{G}$ are constant sheaves, persistent sheaf Laplacians coincide with persistent Laplacians. Since a sheaf cochain complex is constructed through stalks and restriction maps, we expect that persistent sheaf cohomology and persistent sheaf Laplacians contain additional information besides the underlying simplicial complex.

If a simplicial complex $X$ is labeled (each vertex is associated with a quantity $q$), then a sheaf can be constructed as follows.
Let $F: X \to \mathbb{R}$ be a nowhere-zero function. 
We let each stalk be $\mathbb{R}$, 
and for the face relation $[v_0, \dots, v_n] \leqslant [v_0, \dots, v_n, v_{n+1} \dots, v_m]$ (here orientation is not relevant),
the linear morphism $\mathscr{S}([v_0, \dots, v_n] \leqslant [v_0, \dots, v_n, v_{n+1} \dots, v_m])$ is 
the scalar multiplication by
\begin{align*}
     \frac{F([v_0, \dots, v_n])q_{n+1}\cdots q_{m}}{F([v_0, \dots, v_n, v_{n+1}, \dots, v_m])}.
\end{align*}

For a labeled point cloud (a point cloud where each point is associated with a quantity), if we construct a filtration of the point cloud, then for each complex in the filtration we can construct a sheaf as described above. This leads to a filtration of sheaves such as in persistent sheaf cohomology \cite{russold2022persistent} and persistent sheaf Laplacians \cite{wei2025persistent}. The harmonic spectra of PSLs reveal the topological invariants, while the non-harmonic spectra represent geometric information on the data \cite{wei2025persistent,wei2025persistent2}.   
In this work, we use sheaf Laplacians to construct features for individual $C_{\alpha}$ atoms. For a given atom $A$, we first pick a cutoff distance and only consider the nearby $C_{\alpha}$ atoms within the cutoff. Then we choose a radius and build an alpha complex $X$ out of these $C_{\alpha}$ atoms. A cellular sheaf on $X$ is constructed as follows. We denote an atom in $X$ by $v_i$. We assign a label $q_i$ to $v_i$, then, we let each stalk be $\mathbb{R}$. For face relation $v_i \leqslant v_iv_j$, the restriction map is the scalar multiplication by $q_j/r_{ij}$, where $r_{ij}$ is the length of $v_{i}v_{j}$. For face relation $v_iv_j \leqslant v_iv_jv_k$, the restriction map is the scalar multiplication by $q_k/(r_{ik}r_{jk})$. Since we want to distinguish the $C_{\alpha}$ atom $A$ from the other atoms, we let the label of $A$ be 0, and the labels of other nearby $C_{\alpha}$ atoms be 1.  The features are then obtained from the spectra of sheaf Laplacians for this specific $C_{\alpha}$ atom $A$. In this manner, we can construct sheaf Laplacian features for all $C_{\alpha}$ atoms. 

\section{Conclusion}

Protein flexibility is crucial for protein functions, and its prediction is essential for understanding protein properties, protein design, and protein engineering. However, the intrinsic complexity of proteins and their interactions present challenges in understanding protein flexibility. To address this, many effective computational approaches have been developed to predict B-factor values, which reflect protein flexibility. In the literature, a variety of techniques have been proposed, including  NMA \cite{Brooks1983},  GNM \cite{Bahar1998,Bahar1997}, pfFRI \cite{opron2014fast}, ASPH \cite{bramer2020atom}, opFRI \cite{opron2014fast}, and EH \cite{cang2020evolutionary}. 

In this study, we propose a persistent sheaf Laplacian (PSL) model for protein B-factor prediction. Sheaf theory, a branch of algebraic geometry, serves as the foundation for PSL, a novel approach to topological data analysis (TDA). Unlike many global TDA tools, PSL is a localized method that captures the local topology of a point within the data. Similarly to other TDA methods, PSL also provides a multiscale analysis of the system under study.

The multiscale nature of PSL allows it to capture atomic interactions across different distance ranges, enabling a more effective analysis of protein flexibility. This characteristic makes the proposed method superior to traditional approaches, such as GNM, which fail to account for atomic interactions beyond a specific cutoff distance.

For cross-protein prediction, we further enhance the PSL by integrating additional global and local features intrinsic to protein structures and structure determination conditions. This integration enables the blind prediction of protein B-factors, which is particularly valuable for assessing protein flexibility when experimental B-factors are unavailable. The proposed PSL model has been validated using various data sets, demonstrating its effectiveness and robustness in protein flexibility analysis.

\section*{Acknowledgments}
This work was supported in part by NIH grants R01AI164266 and R35GM148196, NSF grant DMS-2052983, MSU Research Foundation, and Bristol-Myers Squibb 65109.

\section*{Data and Code Availability}
Code is available at \href{https://github.com/weixiaoqimath/persistent_sheaf_Laplacians}{https://github.com/weixiaoqimath/persistent\_sheaf\_Laplacians}. \\
Data is available at  \href{https://github.com/fenghon1/MDG_bfactor}{https://github.com/fenghon1/MDG\_bfactor}.

\section*{Author Contributions}
Conception and design: Guo-Wei Wei. Sample preparation and collection of data: Nicole Hayes. Algorithm implementation: Xiaoqi Wei, Hongsong Feng. Analysis and interpretation of data: Nicole Hayes, Guo-Wei Wei. Supervision: Ekaterina Merkurjev, Guo-Wei Wei. Manuscript preparation: Nicole Hayes, Xiaoqi Wei, Hongsong Feng, Ekaterina Merkurjev, Guo-Wei Wei. All authors contributed to the article and approved the submitted version.
 
\section*{Conflict of Interest}
 The authors have no conflicts to disclose.

\bibliography{bib_draft.bib}
\bibliographystyle{unsrt}

\end{document}